\documentclass[aps,prb,showpacs,twocolumn]{revtex4}
\usepackage{graphicx}
\usepackage{amsmath}
\usepackage{amsfonts}
\usepackage{color}

\begin{document}
\title{Ab-initio transport across Bismuth Selenide surface barriers}

\author{Awadhesh Narayan, Ivan Rungger, Andrea Droghetti, and Stefano Sanvito}
\affiliation{School of Physics and CRANN, Trinity College, Dublin 2, Ireland}

\date{\today}

\begin{abstract}
We investigate the effect of potential barriers in the form of step edges on the scattering properties of 
Bi$_2$Se$_3$(111) topological surface states by means of large-scale \textit{ab-initio} transport simulations. 
Our results demonstrate the
suppression of perfect backscattering, while all other scattering processes, which do not entail a complete 
spin and momentum reversal, are allowed. Furthermore, we find that the spin of the surface state develops 
an out of plane component as it traverses the barrier. Our calculations reveal the existence of quasi-bound 
states in the vicinity of the surface barriers, which appear in the form of an enhanced density of states in 
the energy window corresponding to the topological state. For double barriers we demonstrate the formation 
of quantum well states. To complement our first-principles results we construct a two-dimensional low-energy 
effective model and illustrate its shortcomings. Our findings 
are discussed in the context of a number of recent experimental works.
\end{abstract}

\maketitle

%%%%%%%%%%%%%%%%%%%%%%%%%%%%%%%%%%%%%%
\section{Introduction}
%%%%%%%%%%%%%%%%%%%%%%%%%%%%%%%%%%%%%%

Bismuth selenide has emerged as the prototypical topological insulator material due to a single Dirac cone 
in the surface band structure and a relatively large bulk band gap. In 2009, concurrent 
theoretical~\cite{zhang-bi2se3} and experimental~\cite{hasan-bi2se3} works revealed the topological 
insulator phase of Bi$_{2}$Se$_{3}$. Since then many fundamental properties of topological states have 
been demonstrated in this material, which has been called the Hydrogen atom of topological 
insulators.~\cite{spin-pesin,review-kane} 

In recent years, there has been a rapid expansion in the number of scanning tunneling microscope (STM) 
experiments on the Bi$_2$Se$_3(111)$ and the closely related Bi$_2$Te$_3(111)$ surface. Impurities on 
bismuth selenide have been imaged and scattering mediated by bulk states has been 
observed.~\cite{bise-yazdani,bise-xue,bise-kptlnk,bise-kimura} Additionally there have been studies of dopants 
on the bismuth telluride surface~\cite{bite-xue,bite1-kptlnk} and, interestingly, a bound state at a surface step of 
Bi$_2$Te$_3$ has also been found.~\cite{bite2-kptlnk} On the theoretical front, there have been several efforts 
towards modeling scattering of these surface states from perturbation theory by employing Dirac-like model 
Hamiltonians and by imposing symmetry considerations.~\cite{imp-hu, step-bltsky, imp-bltsky} Furthermore, 
a study of robustness of surface states against on-site disorder by employing first-principles calculations was 
also reported.~\cite{guo-disorder} The problem of scattering at a monolayer-bilayer graphene junction has also been investigated.~\cite{nakanishi-graphene}

In this paper, we investigate the effect of step barriers at the Bi$_2$Se$_3$(111) surface on the scattering 
properties of the topological states by means of \textit{ab-initio} transport calculations. We find that, due to 
the spin-polarized helical nature of the surface band, there is no scattering for normal incidence, since a 
reflection would entail a $180^{\circ}$ backscattering. However, as one moves to non-normal incidence, 
scattering is revealed. This is because the spins of the counter-propagating channels are no longer anti-parallel. 
An analysis of the local density of states reveals that the surface barrier strongly affects the spin of the surface 
state, in particular allowing an out of plane spin component, which is negligible in the absence of the barrier. 
In order to compare to our \textit{ab-initio} results we have constructed a potential barrier model based on 
a simple Dirac Hamiltonian for the surface states. This is solved for barriers of various shapes and a comparison 
is made with our first-principles calculations. We note in passing that, although our \textit{ab-initio} 
calculations have been performed for the particular case of bismuth selenide, we expect the same 
qualitative results to also hold for step edges perpendicular to directions without hexagonal warping in Bi$_2$Te$_3$ 
and for other related materials like Bi$_2$Te$_2$S and TlBiSe$_2$. 

Following this introduction, the remaining of the paper is organized as follows. We begin by describing our 
computational methods, in particular we outline the procedure for performing the transport calculations. 
We then study scattering originating from a single surface barrier by analyzing the transmission and the densities 
of states. Intriguingly our calculations reveal a bound state in the vicinity of the barrier, and we study its energy dispersion. From there, we move 
on to construct a low-energy model for the scattering problem, with barrier strength and width extracted from
our \textit{ab-initio} results. Next we look at the analogous problem in the presence of double 
barriers of different lengths. Notably, we find an energy splitting of the bound state when the states at the 
two barriers interact directly as in the case of a short double barrier, as well as when the bound states 
couple with quantum well states formed in the case of a longer double barrier. Finally we summarize 
our results and conclude.

%%%%%%%%%%%%%%%%%%%%%%%%%%%%%%%%%%%%%%
\section{Computational Methods}
%%%%%%%%%%%%%%%%%%%%%%%%%%%%%%%%%%%%%%

%%%%%%%%%%%%%%%%%%%%%%%%%%%%%
\begin{figure}[th]
\begin{center}
  \includegraphics[scale=0.50]{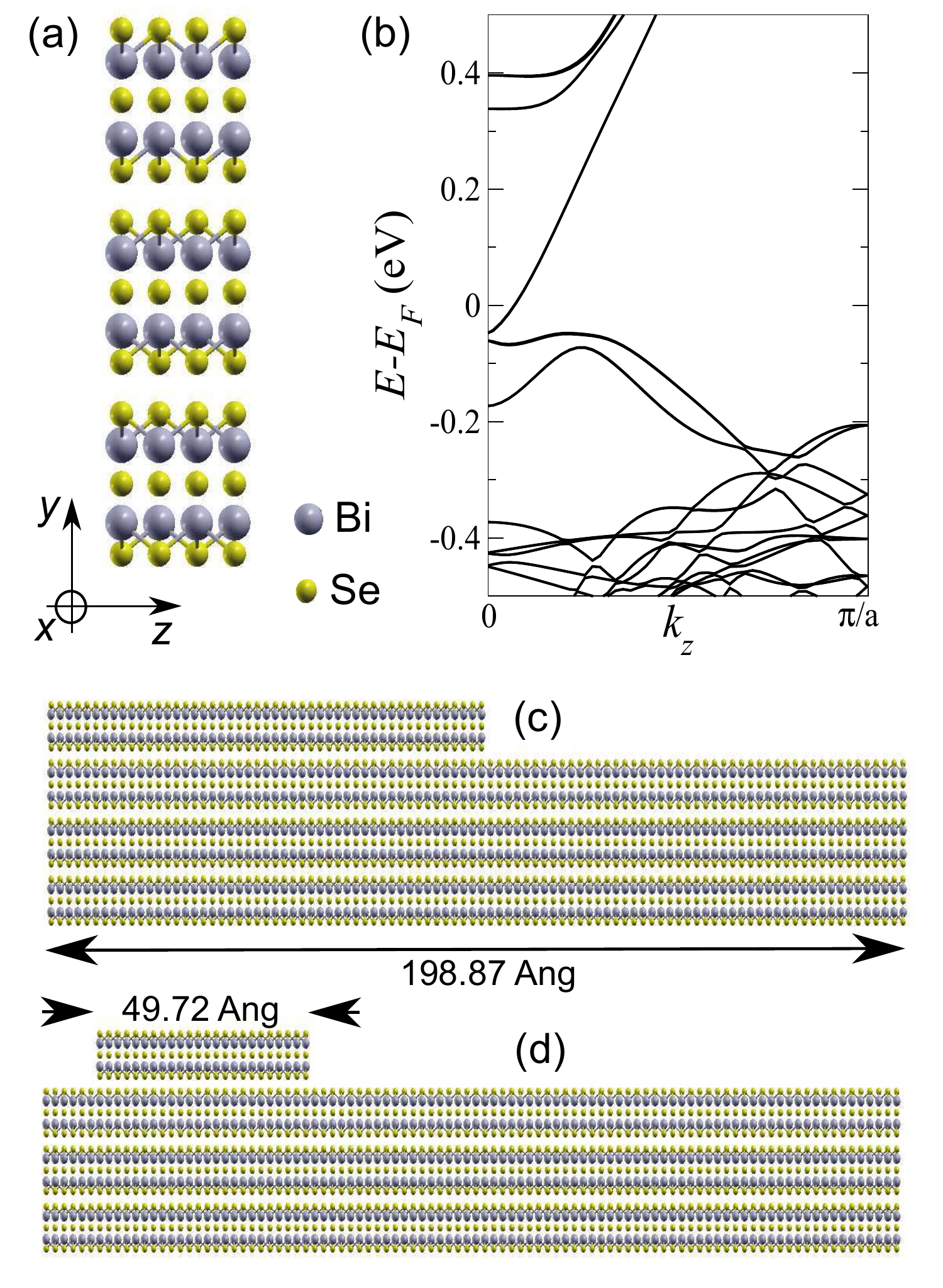}
  \caption{(Color online) (a) The unit cell of the 3-QL slab leads used in the transport calculations, corresponding to a periodic quasi two dimensional system. The 
yellow and purple spheres represent Selenium and Bismuth atoms, respectively. The slab is terminated 
on both sides by Se. (b) The band structure along the transport direction, $z$, is shown at $k_x=0$. The 
surface band in the energy window [-0.05, 0.30]~eV has a helical spin texture, with the spin locked to 
the momentum direction. The transport setup for the scattering problem is shown for (c) a single barrier and 
(d) a double barrier. In both cases we add an extra single quintuple layer high barrier on the 3-QL thick slab. 
Note that the same self-energies for semi-infinite 3-QL leads are attached on the left and right sides of 
the scattering region in (d), while different left and right electrodes corresponding to 4-QL and 3-QL slabs 
are needed in (c).} \label{prelim}
\end{center}
\end{figure}
%%%%%%%%%%%%%%%%%%%%%%%%%%%%%

Our transport calculations have been performed by using the {\sc Smeagol} code, which combines 
the density functional theory (DFT) numerical implementation contained in the {\sc siesta} 
code~\cite{soler-siesta} with the non-equilibrium Green's function method for electron transport. 
Here we briefly outline the calculation procedure and refer the readers to 
References~[\onlinecite{sanvito-smeagol1, sanvito-smeagol2, sanvito-smeagol3}] for a more detailed 
exposition. In {\sc Smeagol} semi-infinite electrodes are attached to a central scattering region by means 
of self-energies. The calculation of the self-energies of the leads is performed by using a singular value 
decomposition-based, robust and efficient algorithm, which overcomes the problems related to recursive 
methods.~\cite{sanvito-smeagol3} The Hamiltonian needed for the algorithm is calculated by using an 
equivalent infinite bulk system. The transport calculation proceeds by using the density matrix and the 
Kohn-Sham Hamiltonian obtained from {\sc siesta}. The self-energies for the leads are added to the 
Hamiltonian of the scattering region, $H$, and the non-equilibrium Green's function is obtained by direct 
inversion
\begin{equation}
 G(E)=\left[E+i0^{+}-H-\Sigma_\mathrm{L}-\Sigma_\mathrm{R}\right]^{-1}\:,
\end{equation}
where $\Sigma_\alpha$ is the self-energy of the left-hand side ($\alpha$=L) and right-hand side 
($\alpha$=R) lead. The charge density is then calculated by integrating the non-equilibrium Green's 
function along a contour in the complex energy plane
\begin{equation}\label{eq2}
 \rho(E)=\frac{1}{2\pi i}\int dE G^{<}(E)\:, 
\end{equation}
where $G^{<}(E)= iG[f_\mathrm{L}\Gamma_\mathrm{L}+f_\mathrm{R}\Gamma_\mathrm{R}]G^{\dagger}$ 
is the lesser Green's function for the transport problem. The Fermi functions for the leads are denoted by 
$f_\alpha$ and $\Gamma_\alpha=(\Sigma_\alpha-\Sigma_\alpha^{\dagger})$ are the broadening matrices. 
In order to perform the contour integral in Eq.~(\ref{eq2}) we use 16 energy points in the complex semicircle, 
16 points along the line parallel to the real axis and 16 poles. The density matrix calculated in Eq.~(\ref{eq2})
is used by {\sc siesta} to re-evaluate the Kohn-Sham Hamiltonian, and such a procedure is iterated until 
self-consistency is obtained. Once convergence is achieved, the relevant quantities like the transmission
function, $T(E)$, and the density of state (DOS), $\mathcal{N}(E)$, are calculated,
\begin{equation}
 T(E)=\mathrm{Tr}[\Gamma_{L}G^{\dagger}\Gamma_{R}G], \quad \mathcal{N}(E)=\frac{1}{2\pi}\mathrm{Tr}[A(E)S]\:,
\end{equation}
where $\mathrm{Tr}$ stands for the trace, $A(E)=i(G-G^{\dagger})$ is the spectral function and $S$ is the 
overlap matrix.

%%%%%%%%%%%%%%%%%%%%%%%%%%%%%
\begin{figure}[ht]
\begin{center}
  \includegraphics[scale=0.40]{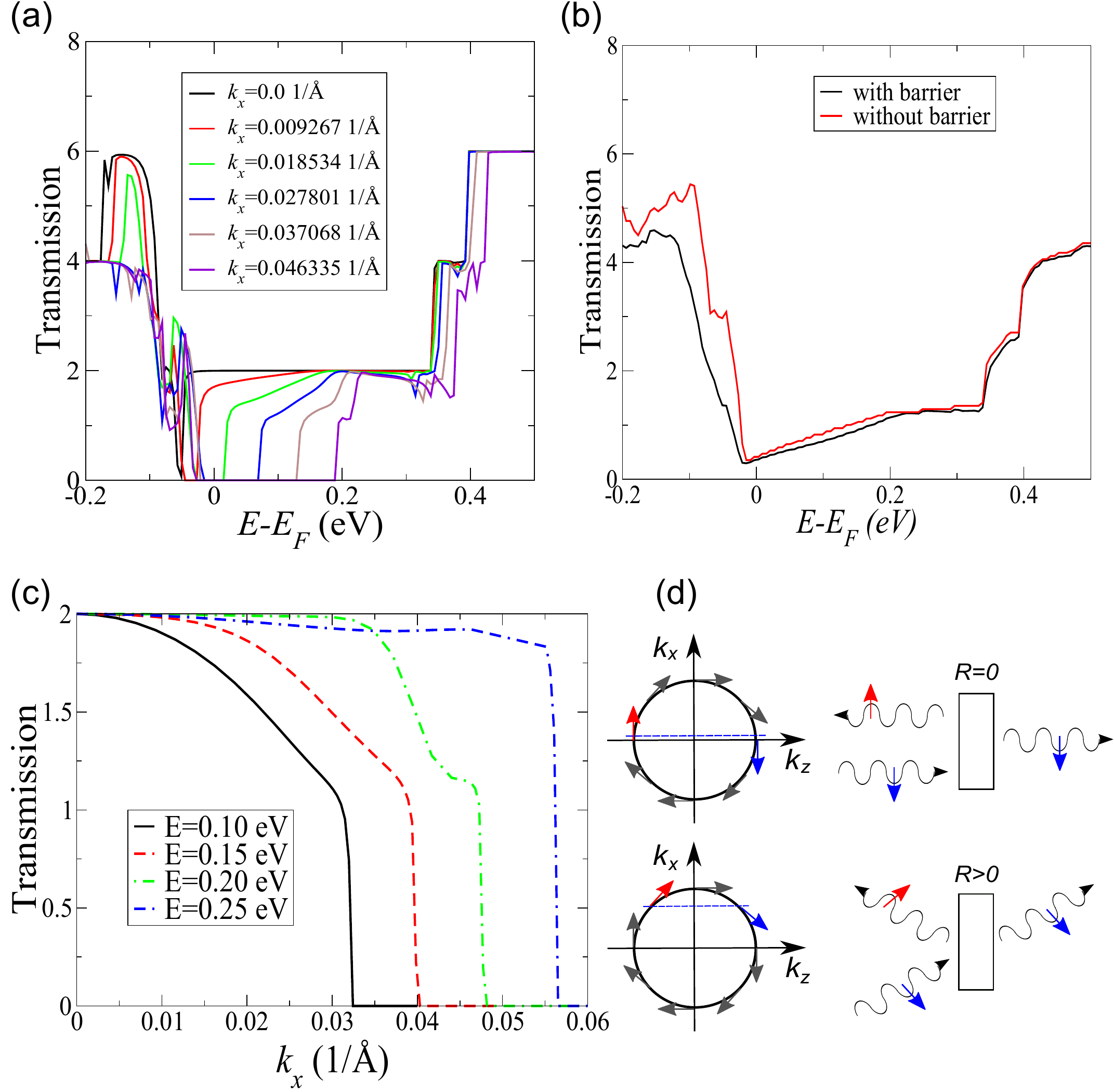}
  \caption{(Color online) (a) Transmission across the surface barrier as a function of energy at different values 
of the $x$ component of the wave-vector, orthogonal to the transport direction. Different curves correspond to 
different $k_{x}$ starting from $k_{x}=0$ up to $k_{x}=0.046335$ \AA{}$^{-1}$, in equal steps of 
0.009267~\AA{}$^{-1}$. Note the perfect transmission at $k_x=0$. At other incidence angles $T$ is reduced. 
(b) The total transmission integrated over $k_{x}$ in the presence (black curve) and absence (red curve) of 
the barrier. (c) The transmission as a function of $k_x$, at different constant energy cuts in the energy region 
of the surface states. Note that $T$ is reduced from $T=2$ at non-zero angle of incidence and with sufficiently 
large $k_x$ drops down towards $T=1$. On further inrease in $k_{x}$, the band edge for the Dirac cone at the bottom surface is reached, and the transmission abruptly goes to zero. Non-zero reflection at the barrier can be explained using the schematic diagram shown in (d). At non-normal incidence there is a finite overlap between the spin projections of the forward and backward moving surface state. Backscattering at angles away from normal incidence is 
present even in the absence of time-reversal symmetry breaking perturbations.} \label{sb_trms}
\end{center}
\end{figure}
%%%%%%%%%%%%%%%%%%%%%%%%%%%%%

In all calculations spin-orbit interaction is included by means of an on-site approximation~\cite{sanvito-soc} and
the Perdew-Burke-Ernzerhof generalized gradient approximation (GGA) to the exchange-correlation functional 
is employed.~\cite{PBE} We have used a double-$\zeta$ polarized basis set and a real space mesh cutoff of 
300~Ryd. For slab calculations a minimum of 25~\AA{} vacuum region has been included to prevent spurious 
interaction between periodic replicas. We use $3\times1\times1$ $k$-point mesh to obtain the self-consistent 
potential (here $x$ is the direction perpendicular to the transport direction in the plane of the slab, $y$ is along 
the slab height and $z$ is the transport direction). When calculating the integrated transmission and DOS we 
use $101$ $k$-points along the $x$ direction. Periodic boundary conditions have been considered in the
plane orthogonal to the transport direction, while using open boundary conditions along the transport 
direction allows us to simulate a single scatterer, which, in this particular case is a surface step.

The unit cell used for the leads is shown in Fig.~\ref{prelim}(a). We note that by adding the self-energies obtained from such quasi two dimensional leads to the Hamiltonian of the scattering region one avoids any finite size effects. It consists of a three quintuple-layers (3-QL) 
thick slab terminated on both sides by Se atoms, as found experimentally. For the slab we use the experimental 
lattice constants. The corresponding band structure is shown in Fig.~\ref{prelim}(b). Note that there is band 
folding as a consequence of the doubling of the Bi$_2$Se$_3$ primitive unit cell. The bandstructure reveals
the Dirac cone and the helical states consistent with earlier studies.~\cite{zhang-bi2se3} It should also be 
noted that there is a small but finite gap (of the order of 0.015 eV) at the $\Gamma$ point in the cone due to 
interaction between the two surfaces at opposite sides of the slab. However, this small gap does not affect 
our analysis of the topological states at higher energies, since the tunneling between the two surfaces is 
negligible as we elaborate in the next section. The transport setup for single and double barrier is shown in Fig.~\ref{prelim}(c) and (d) respectively. 
We consider a single QL-high barrier on a 3-QL thick slab. The step edge is extended along the $\Gamma-M$ 
direction and the transport is along the orthogonal $\Gamma-K$ direction of the primitive Bi$_2$Se$_3$ 
Brillouin zone. The scattering region has a length of 198.87 \AA{}. For the single barrier case the 4-QL 
region extends over about half the length of the scattering region. For the double barrier setup we investigate 
two barrier lengths, where the step extends over a region of 49.72 \AA{} in the shorter case and is 149.16 \AA{} 
for the longer one. Note that such large cells comprising of a few thousand atoms require an accurate order-$N$ algorithm as the one available in
{\sc Smeagol}.~\cite{Mauro}

%%%%%%%%%%%%%%%%%%%%%%%%%%%%%%%%%%%%%%
\section{Scattering from a single barrier}
%%%%%%%%%%%%%%%%%%%%%%%%%%%%%%%%%%%%%%

%%%%%%%%%%%%%%%%%%%%%%%%%%%%%
\begin{figure*}[ht]
\begin{center}
  \includegraphics[scale=0.50]{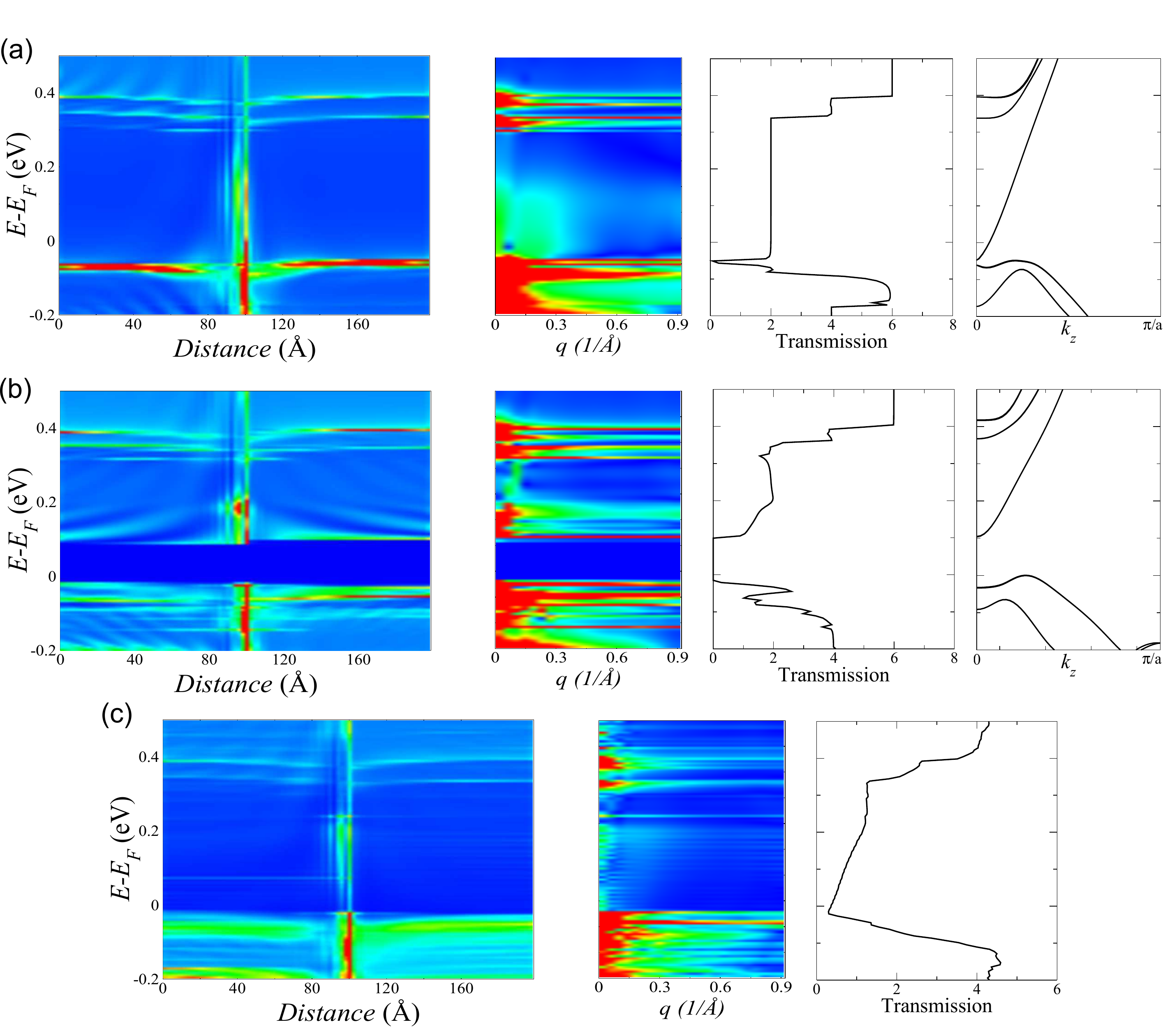}
  \caption{(Color online) The DOS projected on the surface atoms along the scattering region at (a) $k_x=0$, 
(b) $k_x= 0.032$ \AA{}$^{-1}$ and (c) integrated over all $k_x$. At $k_x=0$ there are no oscillations. These
start to emerge at $k_x= 0.032$ \AA{}$^{-1}$ but are not visible in the average. Note in all figures an enhanced 
DOS on the left-hand side of the barrier. The second column of panels show the Fourier transform of the projected
DOS in the flat region adjacent to the barrier, at the corresponding $k_x$. The scattering vector resulting from 
backscattering at non-normal incidence is clearly seen in (b). The average, however, reveals no scattering. Here 
and henceforth warmer colors represent higher and cooler colors indicate lower values, respectively. The third 
column shows the transmission as a function of energy for the three cases. For $k_{x}=0$ and $k_{x}= 0.032$ \AA{}$^{-1}$, 
we also plot the band structure along transport direction for comparison.} \label{sb_pdos}
\end{center}
\end{figure*}
%%%%%%%%%%%%%%%%%%%%%%%%%%%%%

%%%%%%%%%%%%%%%%%%%%%%%%%%%%%
\begin{figure*}[ht]
\begin{center}
  \includegraphics[scale=0.35]{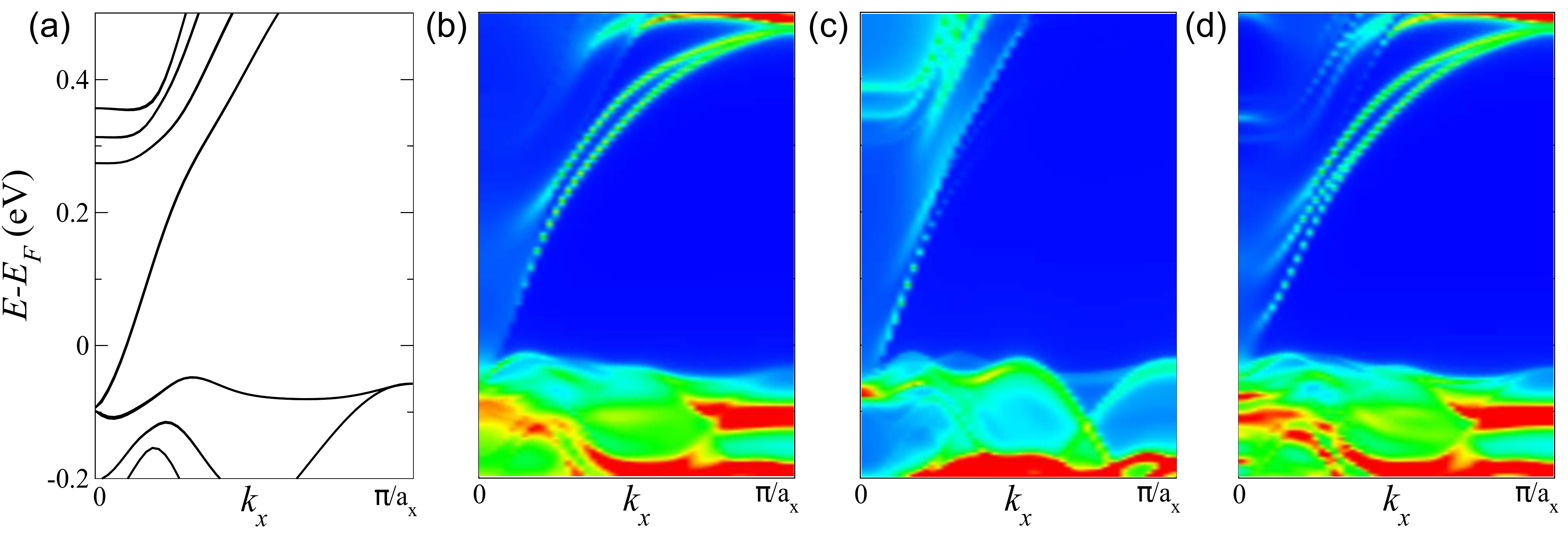}
  \caption{(Color online) The energy dispersion along $k_{x}$ (perpendicular to the transport direction) for (a) perfect periodic system comprising of 4-QL slab, (b) energy dispersion at the single barrier, and (c) 50 \AA{} away from the single barrier. In (b), (c) and (d) color plots show the projected density of states on the atom present at the barrier, an atom 50 \AA{} away from the barrier and the PDOS on the atom at the double barrier (see section V for a discussion). In (b) and (d) note the additional pair of interface states outside the Dirac cone which merge with it around 0.2 eV.} \label{sb_kx}
\end{center}
\end{figure*}
%%%%%%%%%%%%%%%%%%%%%%%%%%%%%

We begin our analysis by looking at the transport across a single surface barrier [see Fig.~\ref{prelim}(c)], 
for which the transmission function is shown in Fig.~\ref{sb_trms}(a) as a function of energy and for different 
values of the ${x}$ component of the wave-vector. At normal incidence ($k_{x}=0$), the surface states are 
perfectly transmitted, $T=2$, due to their helicity. As such, our first-principles calculations confirm Klein 
tunneling.~\cite{katsnelson-graphene} The transmission of bulk states, however, is reduced by the presence of 
the step edge. In contrast, as soon as one moves away from normal incidence, the transmission is no 
longer integer-valued. In particular it dips below $T=2$, indicating substantial scattering. Note that the drop in transmission at $k_{x}=0$ at $E-E_{F}=-0.05$ eV is merely due to the small gap in the band structure due to the finite thickness of the slab. 
Fig.~\ref{sb_trms}(b) shows the total transmission obtained by integrating $T(E,k_x)$ over all angles of 
incidence, namely $T_\mathrm{total}=\frac{1}{\Omega_\mathrm{BZ}}\int_{k_x}T(E,k_x)dk_x$, where 
$\Omega_\mathrm{BZ}$ is the length of the Brillouin zone. Notably $T_\mathrm{total}$ retains the characteristic 
``V-shape'' associated with the linear Dirac cone-like bands, despite the presence of the barrier. Overall we can
conclude that the total transmission in presence of the barrier is quite close to the one for the unperturbed slab
[compare the red and black curves in Fig.~\ref{sb_trms}(c)]. For comparison, we have also performed calculations 
for steps running along the $\Gamma-K$ direction (with transport along $\Gamma-M$). Since the hexagonal 
warping effect, particularly at energies close to the Dirac crossing, is quite small in Bi$_2$Se$_3$, we find 
results, which are very similar to the ones obtained for steps along the $\Gamma-M$ direction. Hence, in the 
rest of this paper we focus our attention on the latter.

At non-normal incidence the spin projections of the surface states counter-propagating at a given edge are 
no longer anti-parallel and thus backscattering becomes allowed, even in the absence of a perturbation that
breaks time-reversal symmetry. We note that, although spin-orbit coupling mixes the spin components, one can still define spin components along different directions by using a projection onto the three Pauli matrices  $\{\sigma_{x},\sigma_{y},\sigma_{z}\}$ and the identity matrix $I$. The situation is schematically illustrated in Fig.~\ref{sb_trms}(d), 
and its consequences are demonstrated in Fig.~\ref{sb_trms}(c), where we plot the transmission across 
the surface barrier as a function of $k_x$ at different energies. Clearly $T(E,k_x)$ is reduced as $k_x$ 
increases, which is expected from argument related to the spin projections of the two counter-propagating 
surface states. At larger incidence angles the transmission tends towards the residual value of one, since 
a perfectly transmitted surface state is present at the opposite side of the slab (no scattering center is present
on the opposite surface). If one increases $k_{x}$ even further, the band edge for the Dirac cone at both surfaces is reached, and the transmission abruptly goes to zero. It can be shown that the maximum scattering amplitude is proportional to 
$\frac{1}{2}(1+\cos\theta)$, where $\theta$ is the angle between the spin directions of the counter-propagating surface 
states.~\cite{hofmann-scatter} Note that at higher energies, the transmission persists at values around the 
unperturbed one, $T=2$, for larger incidence angles. This is because as one moves the Fermi level at
higher energy, the Fermi circle gets larger. Consequently, the same $k_{x}$ corresponds to a smaller incidence angle.

%%%%%%%%%%%%%%%%%%%%%%%%%%%%%
\begin{figure}[ht]
\begin{center}
  \includegraphics[scale=0.50]{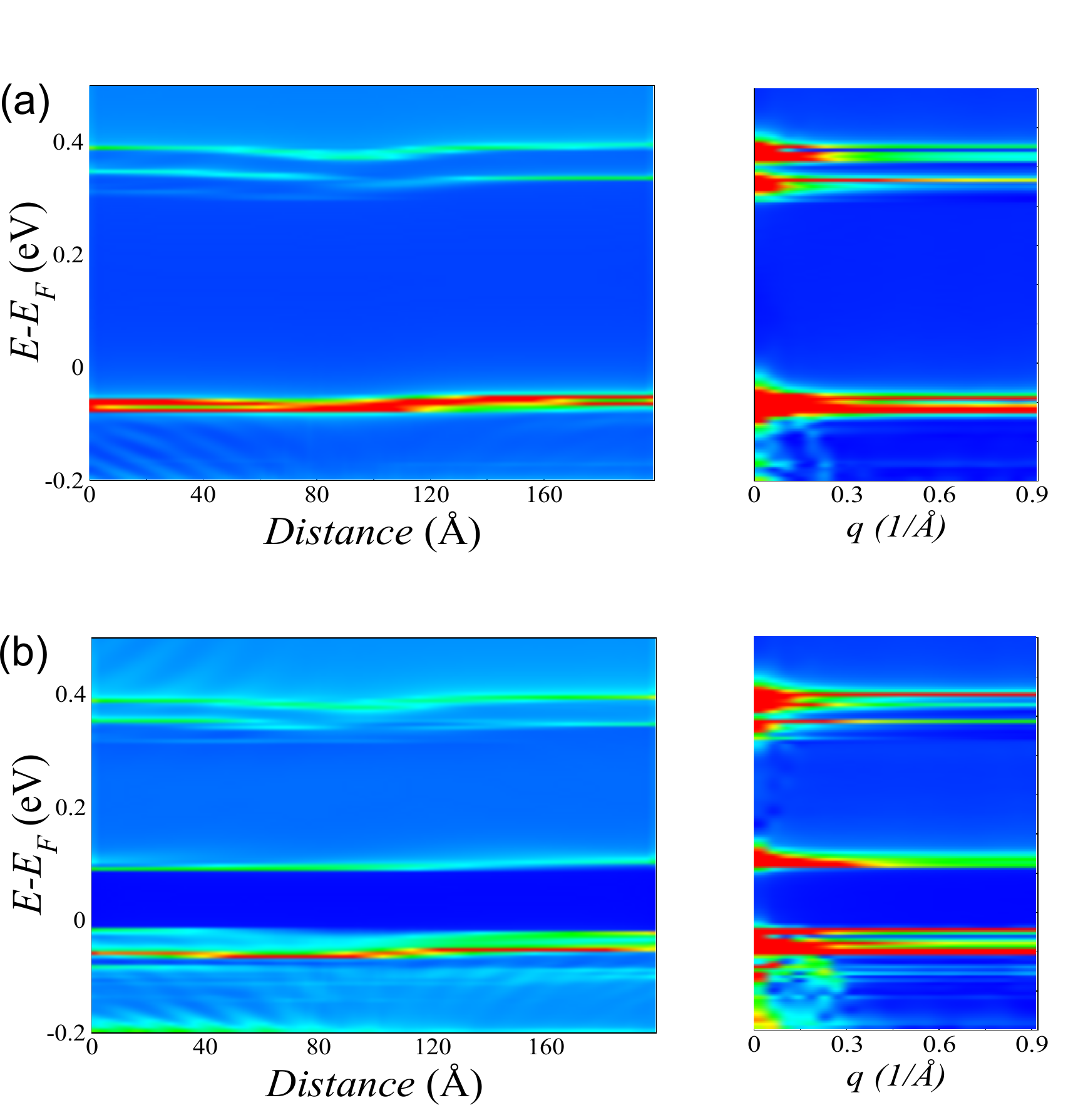}
  \caption{(Color online) The DOS projected on the bottom surface atoms along the scattering region at (a) normal incidence $k_x=0$, and (b) an oblique incidence $k_x= 0.032$ \AA{}$^{-1}$. Note the absence of density oscillations in the bulk enegy gap window, even at non-normal incidence. Panels on the right show the Fourier transform of the projected DOS in the flat region adjacent to the barrier. A comparison with Fig.~\ref{sb_pdos} shows absence of both bound states as well as signature of dominant scattering vectors in the aforementioned energy range. This illustrates a small coupling between the two surfaces of Bi$_2$Se$_3$ slab investigated in this work.} \label{sb_bottom}
\end{center}
\end{figure}
%%%%%%%%%%%%%%%%%%%%%%%%%%%%%

In STM experiments, one measures the oscillations in the electron density in order to study the scattering arising 
from surface modifications, for example from surface steps as studied in Ref.~\onlinecite{eigler-stm}. A Fourier transform of the density yields the characteristic frequencies of its oscillations, i.e, gives the scattering wavevectors, $q=|k_{\mathrm{inc}}-k_{\mathrm{ref}}|$ ($k_{\mathrm{inc}}$ and $k_{\mathrm{ref}}$ are incident and reflected wavevectors, respectively).
In Fig.~\ref{sb_pdos} we plot the density of states projected (PDOS) onto the surface 
atoms along the scattering region. At $k_x=0$ no oscillations in PDOS are seen after reflection from the step 
edge. However, moving away from normal incidence, the above-mentioned oscillations begin to appear. We remind the reader that along the transport direction ($z$) the use of self-energies corresponding to the left-hand side 
and right-hand side electrodes makes the system infinite but non-periodic. Thus no finite size effects, orthogonal to the 
step direction, are seen on the density oscillations. The scattering vectors can be obtained by performing 
a Fourier transform of the DOS along the long flat region adjacent to the barrier. 
At $k_{x}=0$, expectedly there are no prominent scattering processes. 
As one moves to $k_x= 0.032$ \AA{}$^{-1}$, there appears a dominant scattering wave-vector in the Fourier 
transform starting at 0.1~eV and extending upwards in energy, as shown in Fig.~\ref{sb_pdos}(b). This corresponds 
to backscattering at a non-normal incidence angle. Furthermore, this can be mapped to band structure along the
transport direction, where a band starting at the same energy is present. 
The average over $k_x$, however, reveals no scattering on 
this scale, even though there is a clear back scattering at individual $k_x$. In order to accurately resolve the 
small density oscillations above the average, one would need to consider many more $k_{x}$-points in the
calculation. This is computationally prohibitively expensive for the system sizes considered here, and a more 
detailed investigation of the oscillations will be reported elsewhere.~\cite{sanvito-scattering} 
For all three cases we also plot the transmission as a function of energy, for comparison with the surface PDOS.

%%%%%%%%%%%%%%%%%%%%%%%%%%%%%
\begin{figure*}[ht]
\begin{center}
  \includegraphics[scale=0.50]{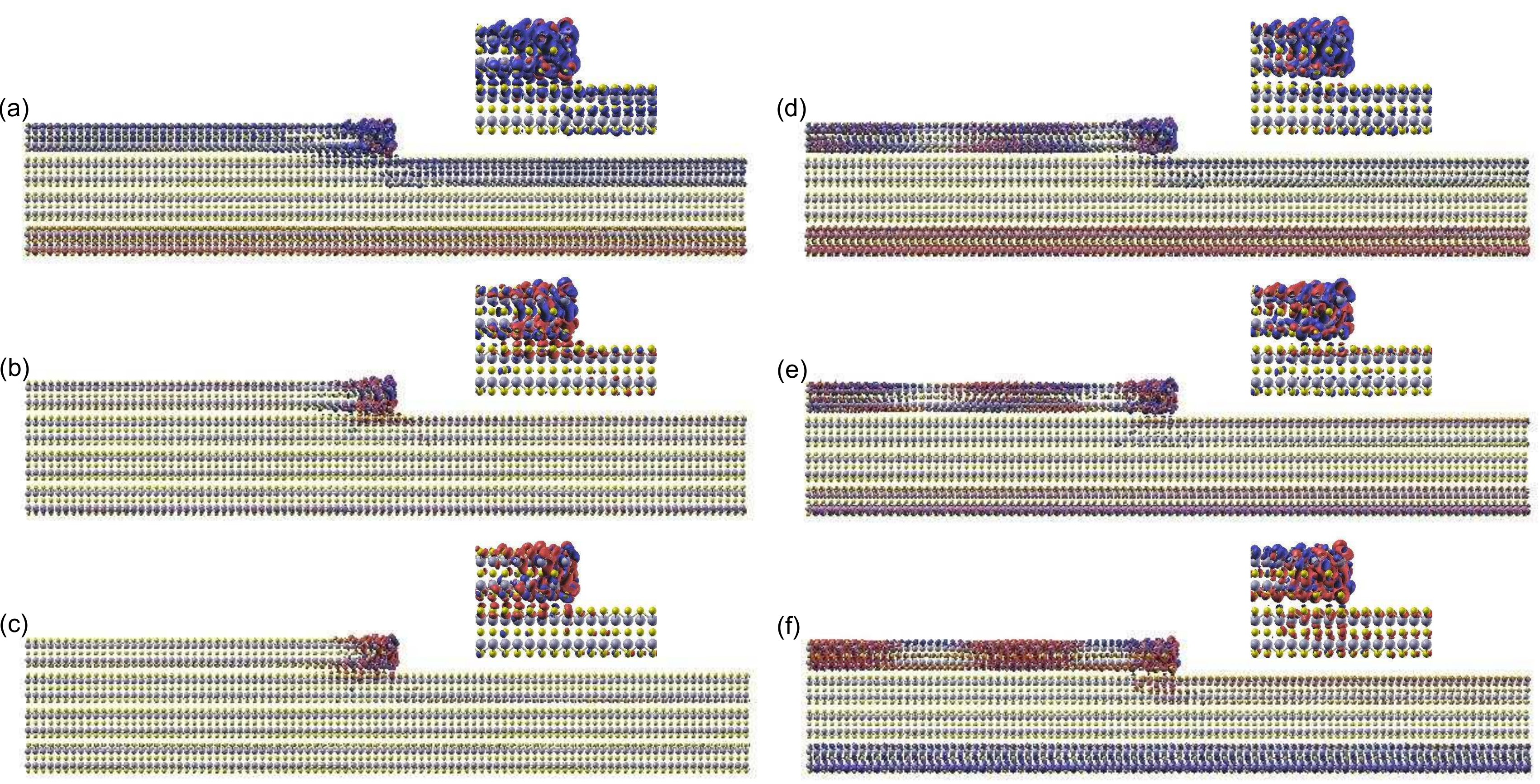}
  \caption{(Color online) The spin-resolved local density of states for states incoming from the left-hand side lead with 
an energy 0.175~eV above the Fermi level. We plot the spin projection along the (a) $x$, (b) $y$ and (c) $z$ 
direction at $k_x=0$. On the right-hand side (d), (e) and (f) are the corresponding plots for $k_x=0.032$ \AA{}$^{-1}$. 
Here red color represents positive values while blue stands for negative ones. Scattering at the step edge even 
at $k_{x}=0$ allows the spin to rotate out of the plane of the slab resulting in finite $y$ and $z$ components, in 
contrast to the unperturbed bottom surface where these are negligible. At non-normal incidence 
($k_x=0.032$ \AA{}$^{-1}$) the $z$ component of the spin-resolved LDOS becomes finite, while the step 
edge introduces a non-zero $y$ component. The insets are zooms around the step edge.  } \label{sb_bubble}
\end{center}
\end{figure*}
%%%%%%%%%%%%%%%%%%%%%%%%%%%%%

Fig.~\ref{sb_pdos} also makes apparent the band bending (of the order of 0.04~eV) introduced by the step. We will show in the next 
section that such band bending close to the step is a crucial ingredient for constructing a scattering model. Far 
enough from the step, however, the PDOS reverts to the unperturbed value within $\sim$40 \AA{}, consistent 
with experimental observation.~\cite{bise-yeh}

In contrast to similar steps on the Sb(111) surface,~\cite{sb-yazdani,sanvito-sbqwell} in Bi$_2$Se$_3$ we find 
bound states close to the step edge and penetrating into the barrier (with an exponentially damped oscillating 
amplitude). These exist over the entire energy window in which the surface states are present. Similar features 
with enhanced DOS have been measured by Alpichshev \textit{et al.}~\cite{bite2-kptlnk} around surface barrier 
at the Bi$_2$Te$_3$ surface. Importantly such a bound state was not ascribed to the warped band structure of 
Bi$_2$Te$_3$. Our results point towards a similar bound state in Bi$_2$Se$_3$ as well. In the experiment, 
no information could be obtained about the DOS on the lower side of the step. Our calculations in fact reveal that 
the state exists only on the higher side of the barrier, and the lower side has no such features. We have also calculated the energy dispersion of this state along the direction perpendicular to the transport. We plot the energy and $k_{x}$ dependence of the PDOS on the Se atom at the barrier [shown in Fig.~\ref{sb_kx}(b)] and on a surface atom 50 \AA{} away from the barrier [Fig.~\ref{sb_kx}(c)], and compare them to the band structure for the perfect periodic system [Fig.~\ref{sb_kx}(a)]. For the atom present at the barrier we find additional pair of states outside the unperturbed Dirac bands, which is consistent with topological band theory. These additional states merge into the Dirac cone at $E-E_{F}\approx 0.2$ eV and produce an enhanced PDOS around that energy. Away from the barrier, however, the PDOS is very similar to that of the unperturbed system, consistent with the pair of bound states being present only close to the barrier. We believe that these predictions of the bound state in Bi$_2$Se$_3$ and its energy dispersion may find verification in future experiments.

By analyzing the PDOS of the atoms at the bottom surface we have checked that significant scattering occurs 
only at the top one, i.e., it is caused by the presence of the step edge and not due to the tunneling back to the 
bottom surface. In Fig. 5 we plot the PDOS on the atoms present at the bottom surface at normal incidence and at a 
representative value of $k_{x}=0.032$ \AA{}$^{-1}$ for oblique incidence. For both cases we can see absence of 
density oscillations in the energy range corresponding to the surface bands. Notably no signature of the bound 
state is also observed. Furthermore, we evaluate the Fourier transform of the PDOS in the flat region next to the 
step and find no features which may be mapped back to the scattering wavevector $q$, in contrast to the case 
of the top surface.

The local density of states (LDOS) associated to electronic states incoming from the left-hand side lead at 
0.175~eV above the Fermi level are shown in Fig.~\ref{sb_bubble}.~\cite{kokalj-xcrysden} These clearly illustrate 
the three-dimensional nature of the path that electrons must traverse while crossing the barrier. The spin projections 
of the LDOS at $k_{x}=0$ and $k_{x}=0.032$ \AA{}$^{-1}$ are shown in the left and right panels, respectively. In 
contrast to pristine bismuth selenide the spins of the helical surface states are no longer confined to the plane of 
the slab. In the vicinity of the barrier they rotate out of the plane (the $y$ component becomes finite). The LDOS at 
the bottom unperturbed surface provide a convenient comparison to the pristine surface, albeit with the spin 
directions reversed. At $k_{x}=0.032$ \AA{}$^{-1}$, the $x$ and $z$ components are dominant for the bottom 
surface, while the step edge introduces a component along the $y$ direction comparable with the other 
two, at the top surface. A zoom close to the step shows a large DOS close to the step edge, which is due to the 
bound state.

%%%%%%%%%%%%%%%%%%%%%%%%%%%%%
\begin{figure}[h]
\begin{center}
  \includegraphics[scale=0.4]{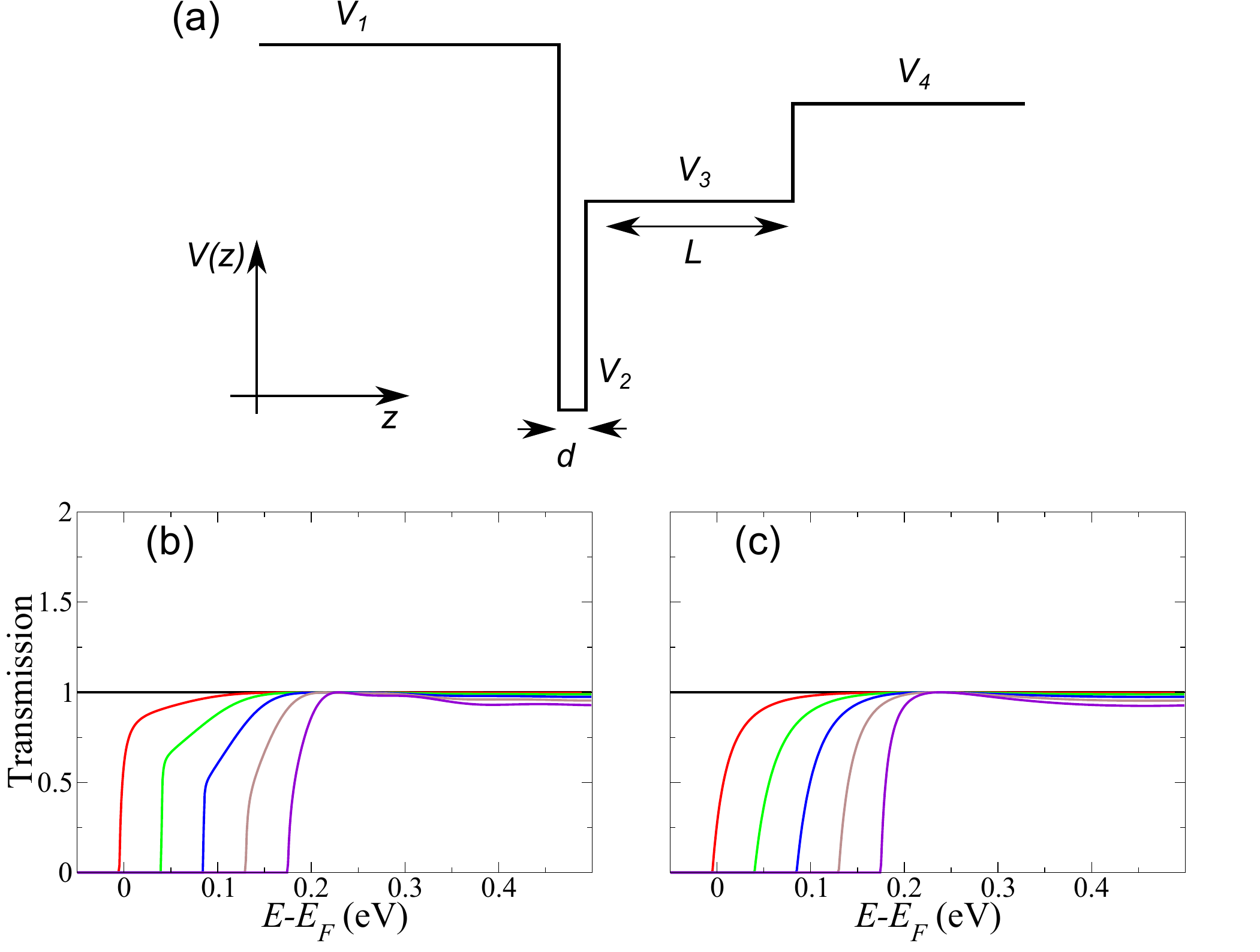}
  \caption{(Color online) (a) Potential profile for the Dirac model. The parameters are fitted to the {\it ab initio} results as following: 
$V_1=V_4=0$, $V_2=-1.17$ eV, $d=20$ \AA{} and $L=60$ \AA{}. The transmission as a function of energy is shown for (b) 
$V_3=-0.02$ eV and (c) $V_3=0.0$ eV. Different curves correspond to different $k_x$ points, with the same definition used in 
Fig.~\ref{sb_trms}(a). } \label{model_trms}
\end{center}
\end{figure}
%%%%%%%%%%%%%%%%%%%%%%%%%%%%%

%%%%%%%%%%%%%%%%%%%%%%%%%%%%%%%%%%%%%%
\section{A low-energy model}
%%%%%%%%%%%%%%%%%%%%%%%%%%%%%%%%%%%%%%

In order to interpret our \textit{ab-initio} results we construct a simple potential barrier model for the scattering problem. The surface states 
are described by a Dirac Hamiltonian~\cite{zhang-bi2se3} 
\begin{equation}
\mathcal{H}= \epsilon_{0} \mathbb{I}_{2\times 2}+ 
\begin{pmatrix}
V(z)& v(k_z-ik_x)\\
v(k_z+ik_x) & V(z)
\end{pmatrix},
\end{equation}
where the potential profile $V(z)$ is shown in Fig.~\ref{model_trms}(a). The values of $\epsilon_{0}=-0.05$ eV and $v=4.58$ eV\AA{}, 
are obtained from our first-principles band structure. Here we consider only the upper part of the cone, i.e., $E=V(z)+\sqrt{k_z^2+k_x^2}$. 
The corresponding eigenstate is given by,
\begin{equation}
 \psi(k_x,k_z)=\frac{1}{\sqrt{2}} \begin{pmatrix} 1 \\ \frac{k_z+ik_x}{\sqrt{k_z^2+k_x^2}} \end{pmatrix} e^{i\mathbf{k}.\mathbf{r}}.
\end{equation}

One can then use the wave-function continuity conditions at the potential steps to solve for the transmission and reflection coefficients in a 
straightforward manner. The potentials in the 4-QL and 3-QL leads, respectively $V_1$ and $V_4$, are nearly identical and are set to zero. 
$V_2$ is the potential associated to the barrier and extends over a length $d$, while $V_3$ is the band bending, which is finite over a 
distance $L$. The calculated transmission curves are plotted in Fig.~\ref{model_trms}(b) for $V_3=-0.02$ eV and in Fig.~\ref{model_trms}(c) 
for $V_3=0$. The shape of the transmission function is much closer to that obtained from the \textit{ab-initio} calculations for finite $V_3=-0.02$ (this value of $V_{3}$ is chosen from our first principles results), as compared to the situation where $V_3=0$. While this comparison does not provide definite evidence of importance of band bending, it serves as an illustration that it is one of the factors which need to be considered while performing a quantitative modeling of step edges on topological insulator surfaces. 
Although it appears that this simplified model can qualitatively reproduce the transmission obtained from first-principles, a more careful analysis shows that it neglects a number of important aspects of the scattering problem. It does not take into account the three-dimensional nature of the barrier, and as a consequence it cannot capture the change in spin orientation of the surface states near the barrier. Moreover it needs as an input the values of the scattering potentials, which an atomistic description is capable of providing, while also capturing the fine details of the scattering process. We also note that several models have been proposed to study topological states on a curved surface. These predict no backscattering at any angle from hyperbolic steps.~\cite{imura-curved1,imura-curved2} Unfortunately these models are not valid for atomic-scale abrupt steps that we have studied in this work.

%%%%%%%%%%%%%%%%%%%%%%%%%%%%%
\begin{figure}[ht]
\begin{center}
  \includegraphics[scale=0.50]{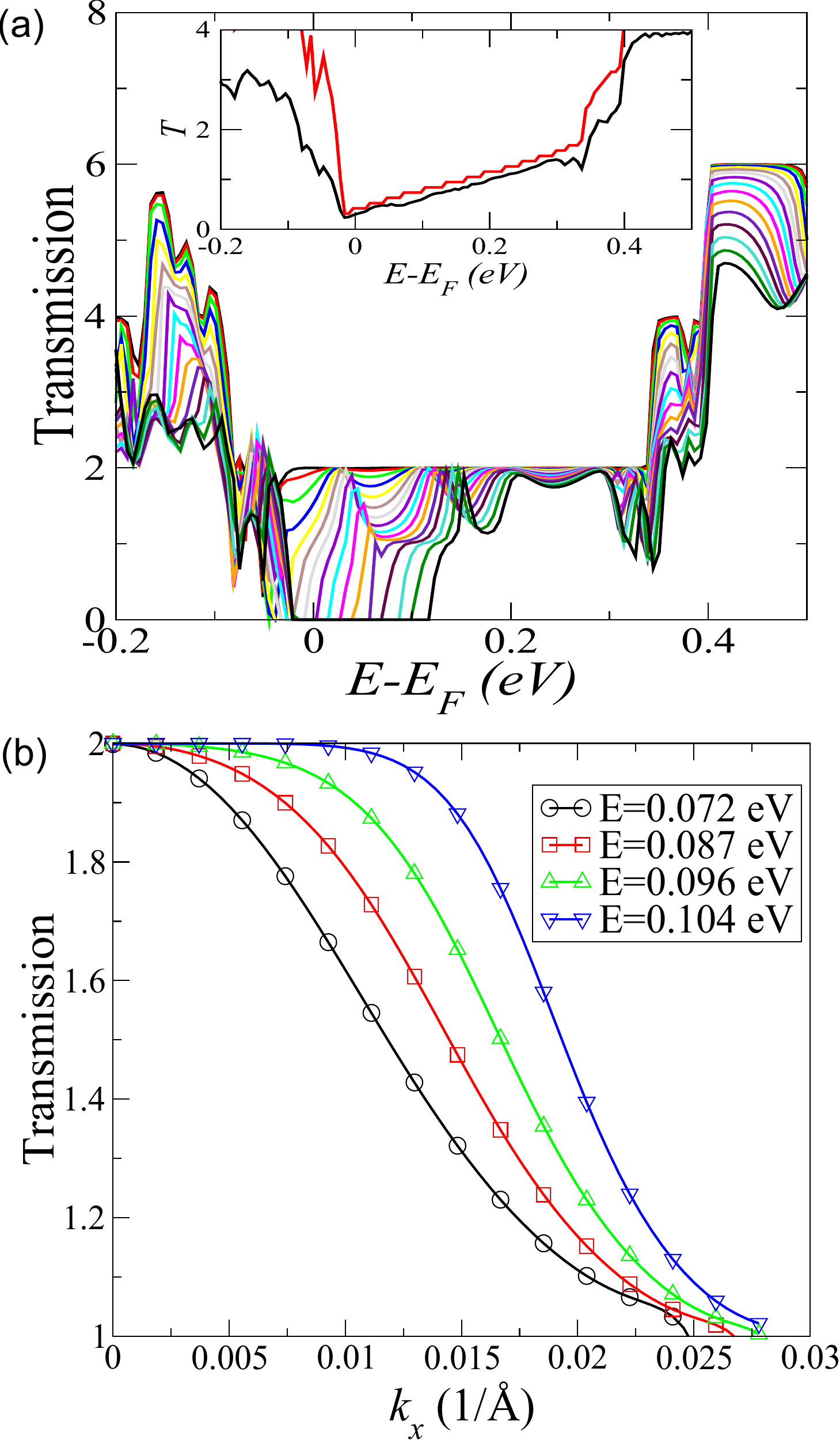}
  \caption{(Color online) (a) The transmission across the double surface barrier at different $k_x$. Similar to the single barrier case, 
$T=2$ for $k_x=0$, but then reduces at other angles of incidence. Note also Fabry-Perot type oscillations in transmission, which are
not present in Fig.~\ref{sb_trms}(a). Different curves correspond to different $k_{x}$ starting from $k_{x}=0$ up to $k_{x}=0.03475$ \AA{}$^{-1}$, in equal steps of 
0.002317~\AA{}$^{-1}$. The integrated transmission obtained with (black curve) and without (red curve) the barriers is plotted in the 
inset. The transmission as a function of $k_x$ at different constant energy cuts is shown in panel (b). These energies are chosen 
away from the resonances.  } \label{db_trms}
\end{center}
\end{figure}
%%%%%%%%%%%%%%%%%%%%%%%%%%%%%

%%%%%%%%%%%%%%%%%%%%%%%%%%%%%%%%%%%%%%
\section{Scattering from double barriers}
%%%%%%%%%%%%%%%%%%%%%%%%%%%%%%%%%%%%%%

%
%%%%%%%%%%%%%%%%%%%%%%%%%%%%%
\begin{figure}[ht]
\begin{center}
  \includegraphics[scale=0.50]{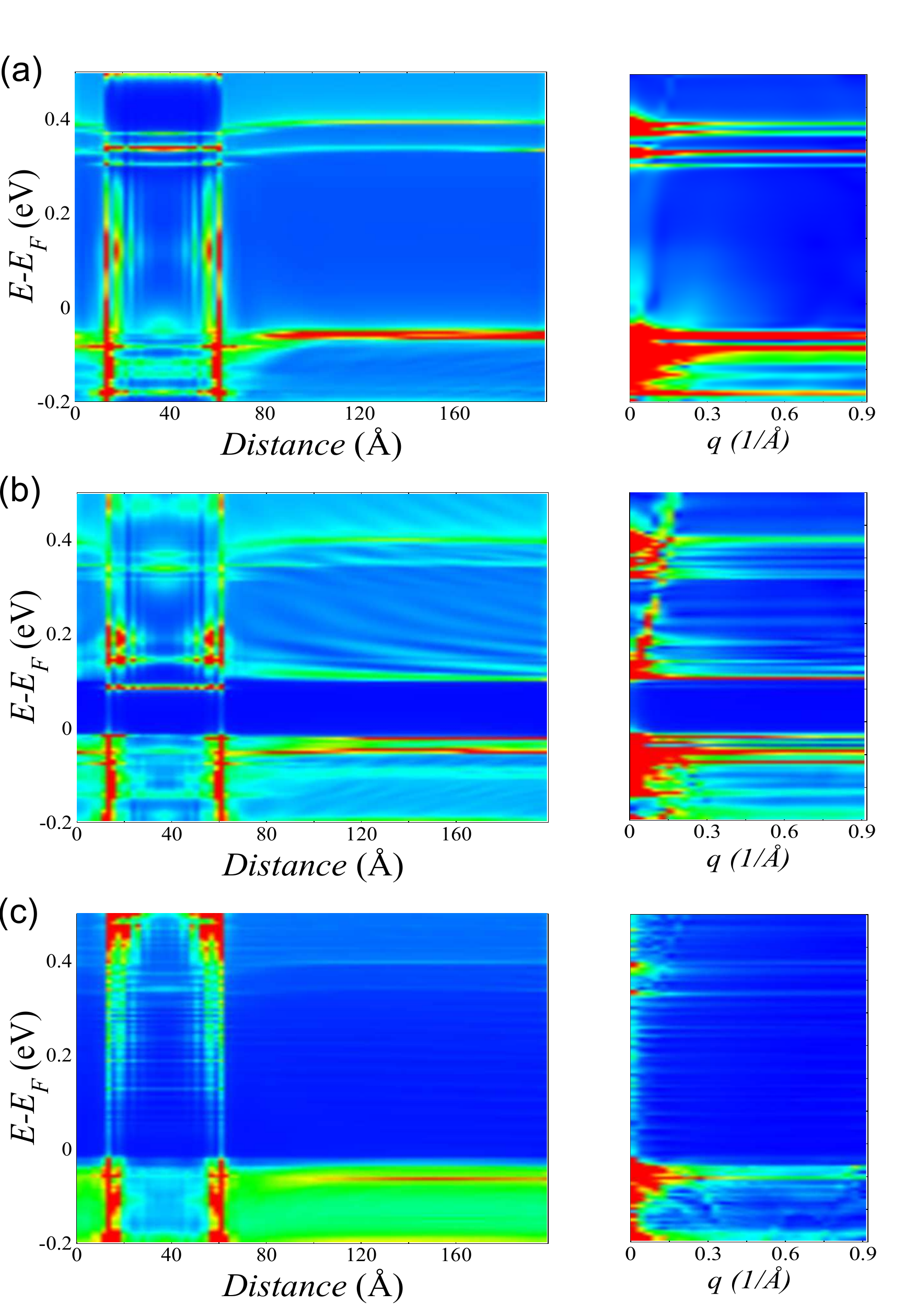}
  \caption{(Color online) DOS projected on the surface atoms along the double barrier scattering region at (a) $k_x=0$, (b) 
$k_x= 0.032$ \AA{}$^{-1}$ and (c) integrated over all $k_x$. Note the absence of density oscillations for $k_x=0$ and for the
integrated DOS. Incidence at finite $k_x$ leads to density oscillations clearly seen in the long flat region adjacent to the barrier 
as shown in (b). The panels on the right-hand side are the corresponding Fourier transforms, which are featureless for (a) and (c), 
while scattering is clearly present in (b). } \label{db_pdos}
\end{center}
\end{figure}
%%%%%%%%%%%%%%%%%%%%%%%%%%%%%

We now analyze the scattering properties of double barrier structures constructed over the Bi$_2$Se$_3$(111) surface. The scattering 
region is shown in Fig.~\ref{prelim}(d) for the shorter surface barrier. This time the scattering structure is connected on both sides to 
two identical semi-infinite leads (3-QL slabs). 
As before, we begin by looking at the transmission across the surface as shown in Fig.~\ref{db_trms}(a). 
Again counter-propagating spin-momentum-locked states yield a perfect transmission at normal incidence. As discussed for the single 
barrier case, at finite $k_x$ the transmission is then reduced. However, in contrast to the previous analysis, there are also resonant energies 
at which the transmission reaches up the value of two, i.e., there is no reflection. At these particular energies the system displays 
Fabry-Perot resonances, which are characteristic of one-dimensional scattering from double potential barriers. In Fig.~\ref{db_trms}(b) we 
plot the transmission as a function of the incident $k_x$ for different energies. Away from the resonances the transmission shows again a 
cosine-like behavior with transmission going down to $T=1$ as the incidence angle increases ($k_x$ gets larger). At even larger $k_{x}$ (not shown) the transmission drops down to zero when the band edge for the Dirac cone is reached at the bottom surface, similar to the case of single barrier.

%%%%%%%%%%%%%%%%%%%%%%%%%%%%%
\begin{figure}[h]
\begin{center}
  \includegraphics[scale=0.65]{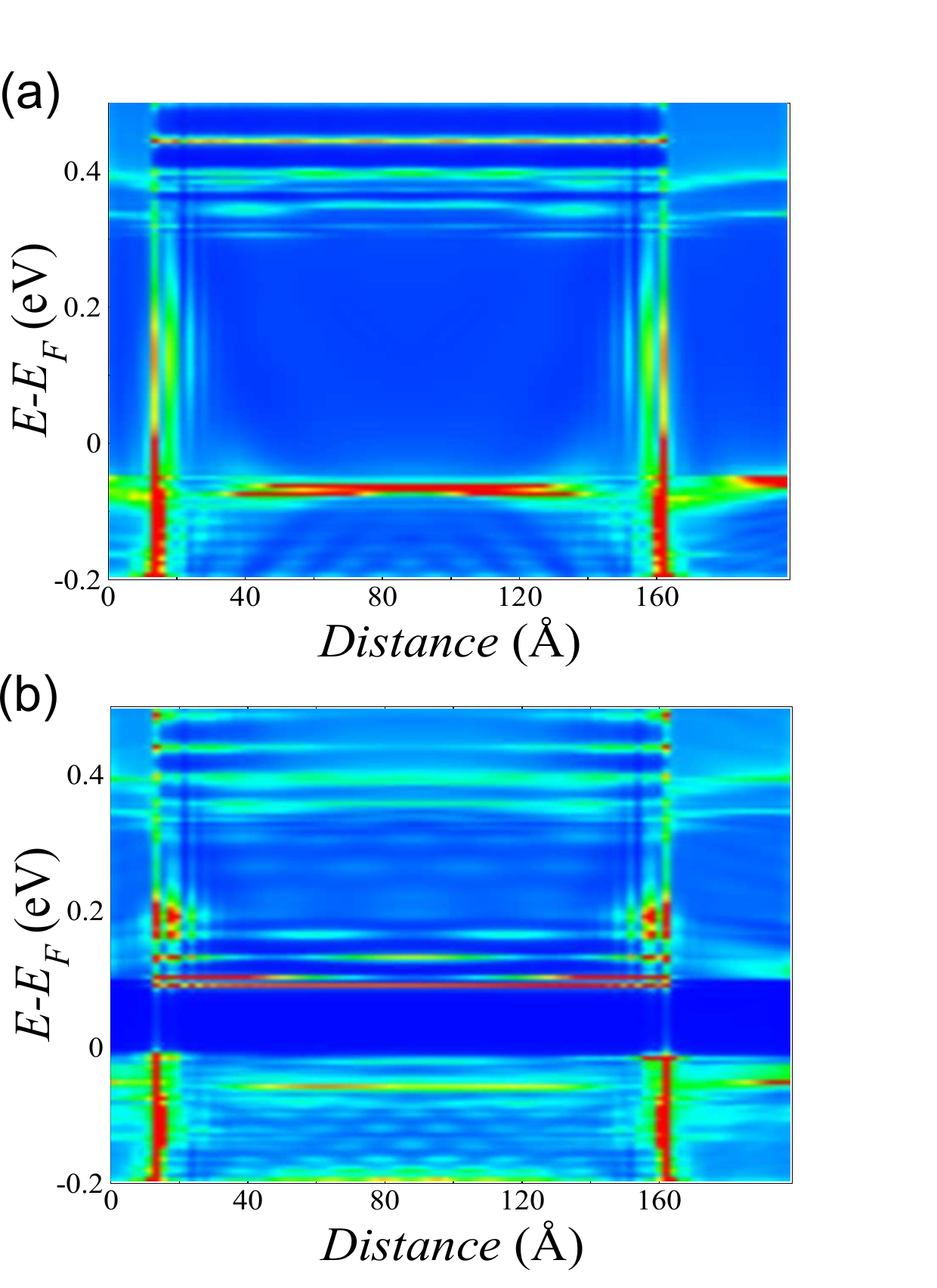}
  \caption{(Color online) PDOS on the surface atoms for a double barrier of length 149.16 \AA{} at (a) $k_x=0$ and 
(b) $k_x=0.032$ \AA{}$^{-1}$. Note the absence of quantum well states in (a). In (b) quantum well states interact with 
the bound state at the two barriers leading to energy splitting of the bound state. } \label{ldb_pdos}
\end{center}
\end{figure}
%%%%%%%%%%%%%%%%%%%%%%%%%%%%%

The $k_x$-resolved and total DOS projected on the surface atoms is plotted in Fig.~\ref{db_pdos}, where the bound state can be clearly 
seen in the 4-QL region extending from 10~\AA{} to 60~\AA{}. The DOS associated to such bound state oscillates and decays towards 
the center of the quantum well defined by the two barriers at the step edges. A band bending similar to that observed for the single barrier 
is also seen for this particular barrier configuration. The interaction between the bound states localised at the two barriers splits them in 
energy, creating alternating high and low DOS as one moves up along the energy axis. Another noticeable feature is a state localized in 
the 4-QL region at around 0.1~eV [see Fig.~\ref{db_pdos}(b)]. This is an additional state in the 4-QL slab, which is decoupled from the 
3-QL leads. The same state is absent in the case of a single barrier produced by a step edge between a 3-QL and a 4-QL semi-infinite
lead. The Fourier transforms of the DOS display similar features as those shown in Fig.~\ref{sb_pdos}. However, in the double barrier
case the resolution is improved over that of the single barrier structure since we now have more atoms along the flat region next to the 
barrier. 

We also study the energy dispersion of the quasi-bound state obtained at the barrier, by calculating the PDOS on the Se atom at the barrier, as a function of $E-E_{F}$ and momentum along the step ($k_{x}$). This is shown in Fig.~\ref{sb_kx}(d), with a comparison to the band structure of the unperturbed periodic system. Apart from the Dirac bands, additional states, dispersing along $k_{x}$ are visible at the interface. These have a dispersion very similar to the case of a single barrier [see Fig.~\ref{sb_kx}(b)]. However, some additional features are seen when this pair of states mixes with the Dirac bands, with an alternating pattern of higher and lower PDOS being visible. This is due to the interaction between the bound states at the two barrier edges. Away from the interface the PDOS and the dispersion reverts to that of the pristine system with only the Dirac bands being present.

Note that for this particular chosen length of the double barrier there are no quantum well states formed inside the 4-QL region. 
However, for a longer barrier the quantum well states appear, as demonstrated by the PDOS on the surface atoms at two different 
$k_x$ for a barrier of length 149.16 \AA{}  (see Fig.~\ref{ldb_pdos}). At normal incidence no quantum well states can be formed in 
the energy window of the surface state, since the two surface states have opposite spin projections leading to no interference. 
In contrast, at  finite $k_x$ quantum well states appear (e.g. a nodeless state at around 0.13~eV and a single-node state at around 
0.16~eV). However, the behavior of these states near the edges of the barrier is different from usual because of the presence of the 
bound state. In fact, these quantum well states interact with the bound states at the edges of the barrier resulting in an energy splitting 
of the bound state. We observe splitting of the bound states in both the short and the long double barrier, in the former case due to 
the interaction between the bound states located at the two edges of the 4-QL region, while in the latter due to the bound state interacting 
with the quantum well state within the barrier.

%%%%%%%%%%%%%%%%%%%%%%%%%%%%%%%%%%%%%%
\section{Summary}
%%%%%%%%%%%%%%%%%%%%%%%%%%%%%%%%%%%%%%

We have used \textit{ab-initio} transport theory to study scattering to both single and double barriers of the topological protected 
states present on a Bi$_2$Se$_3$(111) surface. In particular we have studied the dependence of the transmission on the angle of 
incidence and the electron energy. At normal incidence our first principles approach confirms Klein tunneling. Furthermore, we have 
calculated the density of states projected on the surface atoms and found bound states localised only on the higher side of the barrier. 
Thus our local density of states plots make apparent the three-dimensional nature of the scattering problem, in which the spins of the 
surface states are no longer confined to the plane of the topological insulator slab. We have also constructed a simplified potential 
barrier model using linear Dirac bands to compare with our first principles calculations. Throughout the paper we have placed our 
results in the context of recent experimental works.

%%%%%%%%%%%%%%%%%%%%%%%%%%%%%%%%%%%%%%
\section*{Acknowledgments}
%%%%%%%%%%%%%%%%%%%%%%%%%%%%%%%%%%%%%%

AN thanks the Irish Research Council (IRC) for financial support. IR, AD and SS acknowledge additional financial support by KAUST 
(ACRAB project). Computational resources have been provided by Trinity Centre for High Performance Computing (TCHPC) and 
Irish Centre for High-End Computing (ICHEC).

\end{document}